\documentclass[showpacs,preprintnumbers,amsmath,amssymb,12pt,floatfix,epsfig]{revtex4}
\usepackage{amssymb}
\usepackage{bm}
\usepackage{graphicx}
\newcommand{\be}{\begin{eqnarray}}
\newcommand{\ee}{\end{eqnarray}}

\begin{document}
\draft
\title{High-lying Gamow-Teller excited states in the deformed nuclei, $^{76}\textrm{Ge}$ and $^{82}\textrm{Se}$, by the smearing of the Fermi surface in Deformed Quasi-particle RPA (DQRPA)}

\author{Eunja Ha $^{1)}$ \footnote{ejha@ssu.ac.kr}, K. S. Kim $^{2)}$ and Myung-Ki Cheoun $^{1)}$ \footnote{Corresponding Author: cheoun@ssu.ac.kr}}
\address{1) Department of Physics,
Soongsil University, Seoul 156-743, Korea \\
2)School of Liberal Arts and Science, Korea Aerospace
University, Koyang 412-791, Korea }

\begin{abstract}
With the advent of high analysis technology in detecting the Gamow-Teller (GT) excited states beyond one nucleon emission threshold, the quenching of the GT strength to the Ikeda sum rule seems to be recovered by the high-lying GT states. Moreover, in some nuclei, the stronger GT peaks than any other peaks appear explicitly in the high-lying excited states. We address that these high-lying GT excited states stems from the smearing of the Fermi surface by the increase of the chemical potential owing to the deformation within a framework of the deformed quasi-particle random phase approximation (DQRPA). Detailed mechanism leading to the smearing is discussed, and comparisons to the available experimental data are shown to explain the strong peaks on the high-lying GT excited states in a satisfactory manner.
\end{abstract}

\date{\today}

\maketitle
The deformation in nuclei becomes more important than last decades with the recent development of rare isotope (RI) accelerator facilities, from which one may perform lots of challenging experiments related to the RI nuclei. Most of them are thought to be deformed more or less. They are usually produced in the successive nuclear capture reactions in the cosmos, {\it i.e.} slow- and rapid- process, presumed to be occurred, respectively, at the initial and explosive stage of stellar evolution, such as supernovae (SNe) explosion. Although they decay fractions of a second, they imprint their existence on the nuclear abundances of stars \cite{Haya04}.

Another interesting nuclear process is the neutrino ($\nu$)-process in the SNe explosion which is treated as one of important nuclear processes for the nucleosynthesis, in particular, of the neutron-deficient nuclei, such as $^{138}$La and $^{180}$Ta \cite{Heg,Ch10-2,Ch11-4}. Here $^{180}$Ta is a well known deformed nucleus. Since the neutrinos emitted from a proto-neutron star may have tens of MeV energy high enough to excite the deformed nuclei, one needs to understand more precisely the high-lying excited states beyond one nucleon threshold. Among them, the GT excitations are of great importance because most of the charge exchange reactions are dominated by the GT transition. These high-lying GT excited states are also closely related to the nuclear structure. For example, the GT quenching problem says that the difference of total running sums $S_{\pm}$ for empirical GT ($\pm$) transitions, $S_- - S_+$, is usually quenched compared to the Ikeda sum rule (ISR), $S_- - S_+ = 3 ( N - Z)$.

However, recent experimental GT data on the high-lying states deduced by the more energetic projectiles shed a new light on the GT states located above one nucleon threshold, whose contributions are thought to enable us to explain the quenching problem with the high-lying GT excited states through the multi-particle and multi-hole configuration mixing \cite{Yako05}. Of course, the contributions from the $\Delta$ excitation and from the two-body current may be still effective for the quenching problem.

Conventional approach to understand the nuclear
structure is based on the spherical symmetry. In order to describe
neutron-rich nuclei and their relevant nuclear reactions occurred in the nuclear processes, one needs to
develop theoretical formalism including explicitly the deformation
\cite{simkovic,saleh}. Ref. \cite{simkovic} exploited the Nilsson basis to the deformed quasi-particle random phase
approximation (DQRPA). But the two-body interaction was derived from the effective separable force.
The realistic two-body interaction is firstly considered at Ref. \cite{saleh}
only with neutron-neutron (nn) and proton-proton (pp) pairing correlations which have only isospin T = 1 and J = 0 interaction.
But, to properly describe the deformed nuclei, the T= 0 and J = 1 pairing should be also taken into account
because it may easily lead to the deformation by the J = 1 pairing.

In this work, we extend our previous QRPA based on the spherical symmetry \cite{Ch93},
which has been exploited as a useful framework for describing the nuclear reactions sensitive
on the nuclear structure of medium-heavy and heavy nuclei \cite{ch10}. Since our DQRPA formalism is fully discussed at Ref. \cite{Ha2012}, we briefly summarize the DQRPA used in this work. Starting from the deformed Wood Saxon potential \cite{Hama0407}, we transform a physical state given by the diagonalization of total Hamiltonian in the Nilsson basis into the spherical basis, in which one may perform more easily theoretical calculations.

In a cylindrical coordinate, eigenfunctions of a single particle state and its time-reversed state denoted as $\alpha$ and ${\bar \alpha}$ in deformed Woods-Saxon potential are expressed as follows
\begin{eqnarray}\label{eq:Nil}
&&|\alpha \rho_{\alpha}=+1> =\sum_{N n_z} [ b_{N n_z \Omega_{\alpha}}^{(+)}~
|N, n_z, \Lambda_{\alpha}, \Omega_{\alpha}= \Lambda_{\alpha}+ 1/2 >  \nonumber \\
&&+  b_{N n_z \Omega_{\alpha}}^{(-)}|N, n_z, \Lambda_{\alpha}+1, \Omega_{\alpha}= \Lambda_{\alpha}+ 1-1/2 >],
\end{eqnarray}
where $N=n_\perp + n_z$ ($n_\perp =2n_\rho + \Lambda_{\alpha} $) is a
major shell number, and $n_z$ and $n_\rho$ are numbers of nodes of the physical state on the deformed harmonic oscillator wave functions in $z$ and
$\rho$ directions, respectively. $\Lambda_{\alpha} (\Omega_{\alpha})$ is the
projection of the orbital (total) angular momentum onto the
nuclear symmetric axis $z$. The coefficients $b^{(+/-)}_{N n_z \Omega_{\alpha}}$ are obtained by the eigenvalue equation of the total Hamiltonian in the Nilsson basis.

The deformed harmonic oscillator wave function, $|N n_z \Lambda_{\alpha} \Sigma ~( = \pm 1/2) >$ in Eq. (\ref{eq:Nil})
can be expanded in terms of the spherical harmonic oscillator wave function $|N_0 l \Lambda \Sigma > $ as follows
\begin{equation}\label{eq:trans}
|N n_z \Lambda_{\alpha}~\Sigma > =
\sum_{N_0,l }
A_{N n_z \Lambda}^{N_0 l, ~n_{r}={N_0 -l \over 2}}~ \sum_{j} C_{l \Lambda_{\alpha}{1 \over 2}\Sigma}^{j \Omega_{\alpha}}
|N_0 l j ~\Omega_{\alpha}>,
\end{equation}
with the Clebsch-Gordan coefficient $ C_{l \Lambda_{\alpha} { 1 \over 2} \Sigma}^{j \Omega_{\alpha}}$ and the spatial overlap
integral $A_{N n_z \Lambda}^{N_0 l} =<N_0 l \Lambda|N n_z \Lambda
>$ numerically calculated in the spherical coordinate
system. Therefore, the expansion can be simply written as $
|\alpha \Omega_{\alpha}> =\sum_{a} B_{a}^{\alpha}~|a \Omega_{\alpha} >,
$ with the expansion coefficient summarized as
$ B_{a}^{\alpha} = \sum_{N n_z \Sigma} C_{l \Lambda { 1 \over 2} \Sigma}^{j \Omega_{\alpha}}
A_{N n_z \Lambda}^{N_0 l}~b_{N n_z \Sigma}$.

The Deformed BCS equation is solved by using {\it ab initio} Brueckner G-matrix calculated from the realistic Bonn CD potential
for the nucleon-nucleon interaction. For example, the pairing potentials $\Delta_{p}$ is calculated as
\begin{eqnarray} \label{eq:gap}
&&\Delta_{\alpha p \bar{\alpha}p} = - {1 \over 2} {1 \over (2 j_a +1)^{1/2}}
\sum_{J, c }g_{{pair}}^{p} F_{\alpha a \alpha a}^{J0} F_{\gamma c \gamma c}^{J0}
 \nonumber \\
&&G(aacc,J)(2 j_c +1)^{1/2} (u_{1p_c}^* v_{1p_c} + u_{2p_c}^* v_{2p_c}) ~,
\end{eqnarray}
where $F_{ \alpha a  {\bar \beta b
}}^{JK'}=B_{a}^{\alpha}~B_{b}^{\beta} ~C^{JK'}_{j_{\alpha}
\Omega_{\alpha} j_{\beta}\Omega_{\beta}}(K'=\Omega_{\alpha}+\Omega_{\beta})$ is designed for the transformation
to the deformed basis of the G-matrix. Here $K'$, which is a projection number
of the total angular momentum $J$ onto the $z$ axis, is selected $K'=0$ at the BCS stage
because we consider the pairings of the quasi-particles at $\alpha$ and ${\bar\alpha}$ states. In order to renormalize the G-matrix, strength parameters,
$g_{{pair}}^{p}$, $g_{{pair}}^{n}$ and $g_{{pair}}^{pn}$ are multiplied to the G-matrix to reproduce empirical pairing gaps \cite{Ch93}.
The ${\beta}^{\pm}$ decay operator, ${\hat\beta }_{1 \mu}^{\pm}$, is defined in the intrinsic frame as
\begin{equation} \label{eq:btop}
{\hat\beta }_{1 \mu}^{-}  = \sum_{\alpha_p \rho_\alpha \beta_n \rho_\beta }
< \alpha_{p}\rho_{\alpha} |\tau^{+} \sigma_K | \beta_{n}\rho_{\beta}>  c_p^{\dagger} {\tilde c}_n .~
\end{equation}
The $\beta^{\pm}$ transition amplitudes from the ground state of an initial nucleus
to the excited state are expressed by
\begin{eqnarray}
&&< 1(K),m | {\hat\beta}_{K }^- | ~QRPA >   \nonumber \\
&&= \sum_{\alpha \alpha''\rho_{\alpha} \beta \beta''\rho_{\beta}}{\cal N}_{\alpha \alpha''\rho_{\alpha}
 \beta \beta''\rho_{\beta} }
 < \alpha \alpha''p \rho_{\alpha}|  \sigma_K | \beta \beta''n \rho_{\beta}> \nonumber \\
&& [ u_{p \alpha \alpha''} v_{n \beta \beta''} X_{\alpha \alpha''  \beta \beta'',K} +
v_{p \alpha \alpha''} u_{n \beta \beta''} Y_{\alpha \alpha'' \beta \beta'',K}  ] ~,
\end{eqnarray}
where $|~QRPA >$ denotes the correlated QRPA ground state in the intrinsic frame with the nomalization factor given as $ {\cal N}_{\alpha \alpha'' \beta
 \beta''} (J) = \sqrt{ 1 - \delta_{\alpha \beta} \delta_{\alpha'' \beta''} (-1)^{J + T} }/
 (1 + \delta_{\alpha \beta} \delta_{\alpha'' \beta''}). $ The Wigner functions disappeared by using the orthogonality of two Wigner functions from the operator and the excited state, respectively.
To compare to the experimental data, the GT($\pm$) strength functions
\begin{equation}
B_{GT}^{\pm}(m)= \sum_{K=0,\pm 1} | < 1(K),m || {\hat \beta}_{K }^{\pm} || ~QRPA > |^2
\end{equation}
and their running sums $S_{GT}^{\pm} = \Sigma_m B_{GT}^{\pm}(m)$ for $^{76}$Ge and $^{82}$Se are enumerated as a ratio of the $S_{GT}^- - S_{GT}^+$ to the ISR value $3 (N-Z)$ calculated with the closure relation (ISR I) and without it (ISR II).
\begin{table}
\caption[bb]{Deformation parameters $\beta_2$,
pairing strength parameters $g_{{pair}}^{{p}}(g_{{pair}}^{{n}})$, and theoretical pairing gaps $\Delta^{{p,n}}_{{th}}$ used in this work.
The ISR I and II in the last column are given as $ ( S_{GT}^- - S_{GT}^+) / 3(N-Z)$.
The particle-particle (particle-hole) strength parameters are taken as $g_{{pp}}=1.0 (g_{{ph}}=1.0)$ for two nuclei.
Empirical pairing gaps, $\Delta^{{p}}_{{em}}$ ($\Delta^{{n}}_{{em}}$)=1.561 (1.535) and 1.409 (1.544) Mev, are exploited for $^{76}\textrm{Ge}$ and $^{82}\textrm{Se}$, respectively. }
\setlength{\tabcolsep}{2.0 mm}
\begin{tabular}{cccccc}\hline
 A & $\beta_2$& $g_{{pair}}^{{p}}(g_{{pair}}^{{n}})$ &  $\Delta^{{p}}_{{th}}$& $\Delta^{{n}}_{{th}}$ &ISR I, II($\%$)
  \\ \hline \hline
 ${}^{76}$Ge  &  0.1    & 1.10(1.37)        & 1.575   & 1.538       & 97.04, 106.03 \\
            &  0.2    &  1.17(1.41)       & 1.578   & 1.540         & 96.94, 102.40 \\
            &  0.35    &  1.52(3.19)       & 1.562   &  1.545       & 96.48, 99.54  \\ \hline
            &  0.1    &  1.01(1.49)       & 1.502   &  1.579        & 96.51, 106.45 \\
${}^{82}$Se & 0.2     &  1.11(3.49)       & 1.419   &  1.547        & 95.62, 107.55 \\
            & -0.1    &  1.02(1.53)       & 1.410   &  1.564        & 96.58, 104.89 \\
            & -0.2    &  1.01(1.64)       & 1.568   &  1.549        & 96.57, 109.69 \\ \hline
 \end{tabular}
\label{tab:result1}
\end{table}

The single particle states are used up to $4 \hbar \omega$ for two nuclei, in the spherical limit.
Since the GT strength distribution turns out to rely on the deformation parameter, $\beta_2$ \cite{Hama0407},
we exploited the deformation parameter, $|\beta_2| \le 0.35$, as the default values.
Corresponding values of the deformation parameter $\beta_2$, the renormalized strengths
$g_{{pair}}^{n}$ and $g_{{pair}}^{p}$, and theoretical and empirical pairing gaps are tabulated in Table I.
The ISR I and II, which are deviated 8\% maximally from the ISR, seems to be reasonable more or less, if we remind that the particle model spaces spanned by the deformed basis are finite less than 4 $\hbar \omega$ and two-body currents including $\Delta$-excitation are not taken into account in this work.

An interesting and important point to be noticed is that the $g_{pair}^{n}$ values are abnormally deviated from 1 for some prolate $\beta_2$ values. This turns out to results from the
wide smearing of the Fermi surface in deformed nuclei, which makes the $uv$ coefficients for the paring gaps in Eq.(3) small around the Fermi surface and enlarges the $g_{pair}^n$ to fit the empirical pairing gap.

This tendency is explicitly revealed in the occupation probabilities of $ ^{82}$Se in Fig.1, which shows a wide smearing, {\it i.e.} significant change of the occupation probabilities of the particles around the Fermi surface by the increase of the deformation parameter $\beta_2$. Since the increase of the prolate deformation usually raises the Fermi surface energy, some states above the Fermi surface burrowed below or around the Fermi surface. For example, (4 3 1 1/2), (4 2 2 3/2) and (4 1 1 3/2) states in $^{82}\textrm{Se}$ are such cases. This gives rise to the wide smearing. Such wide smearing of the Fermi surface is also found for the protons and neutrons in $^{76}$Ge, as shown in Fig.2.
\begin{figure}
\includegraphics[width=1.0\linewidth]{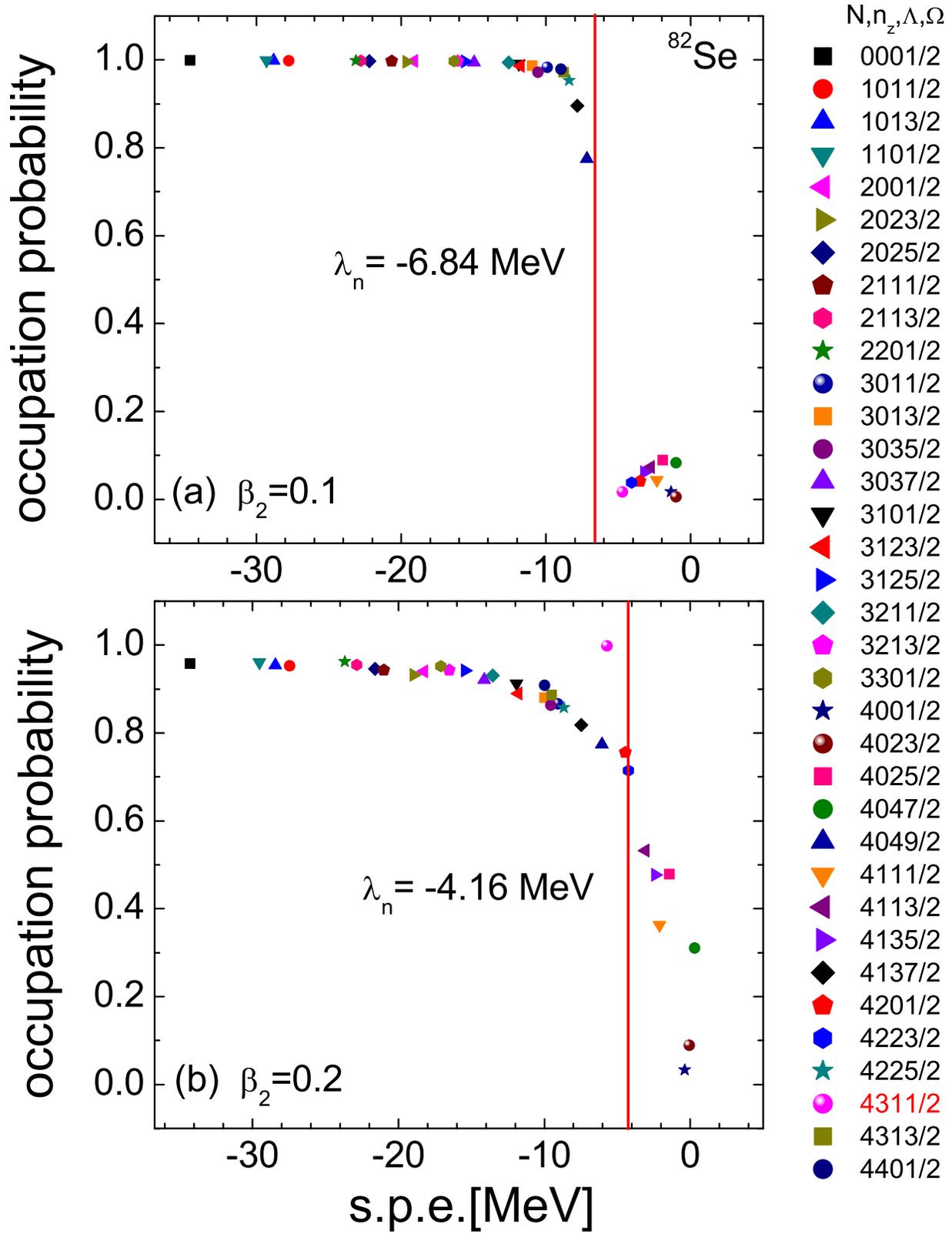}
\caption{(Color online) Occupation probabilities of the neutrons in $ ^{82}$Se
as a function of the single particle energy given by Nilsson basis for two deformation parameters $\beta_2$. With the increase of $\beta_2$ from 0.1 to 0.2, the Fermi energy $\lambda_n$ is increased with the wider smearing of the Fermi surface.}
\label{fig1}
\end{figure}
\begin{figure}
\includegraphics[width=1.0\linewidth]{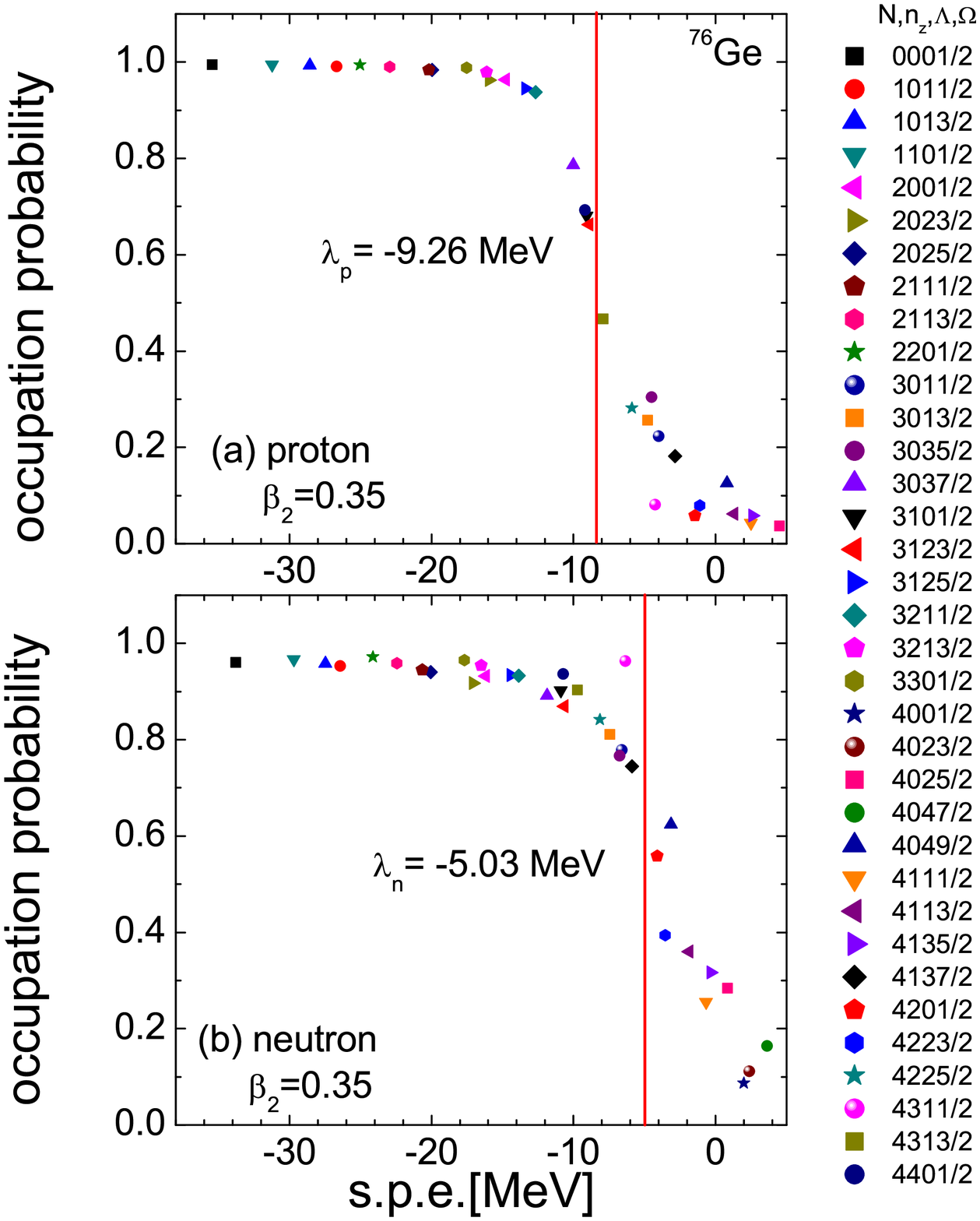}
\caption{(Color online) Same as Fig.1 but with $ ^{76}$Ge. For comparison, results for protons are shown in the left panel. }
\label{fig2}
\end{figure}

Detailed changes of single particle states by the $\beta_2$ deformation leading to the smearing are presented in Figs.3 and 4. Some states above the Fermi surface are reallocated below or around the surface, because the single particle state energies adopted from the deformed Woods Saxon potential depend on the parameter $\beta_2$. The deformation of nuclei may be conjectured to come from macroscopic phenomena, for example, the core polarization, the high spin states and so on. Microscopic reasons may be traced to the tensor force in the nucleon-nucleon interaction, which is known to account for the shell evolution according to the recent shell model calculations \cite{Otsuka05,Otsuka10}. For example, T = 0, J = 1 pairing, which is associated with the $^{3}S_0$ tensor force, may lead to the deformation compared to the T = 1, J = 0 pairing.

Therefore, the deformation parameter adopted in this work may include implicitly and effectively such effects, because the single particle states from the deformed Wood Saxon potential show strong dependence on the $\beta_2$, as shown in Figs. 3 and 4 \cite{Nilsson}.
\begin{figure}
\includegraphics[width=1.0\linewidth]{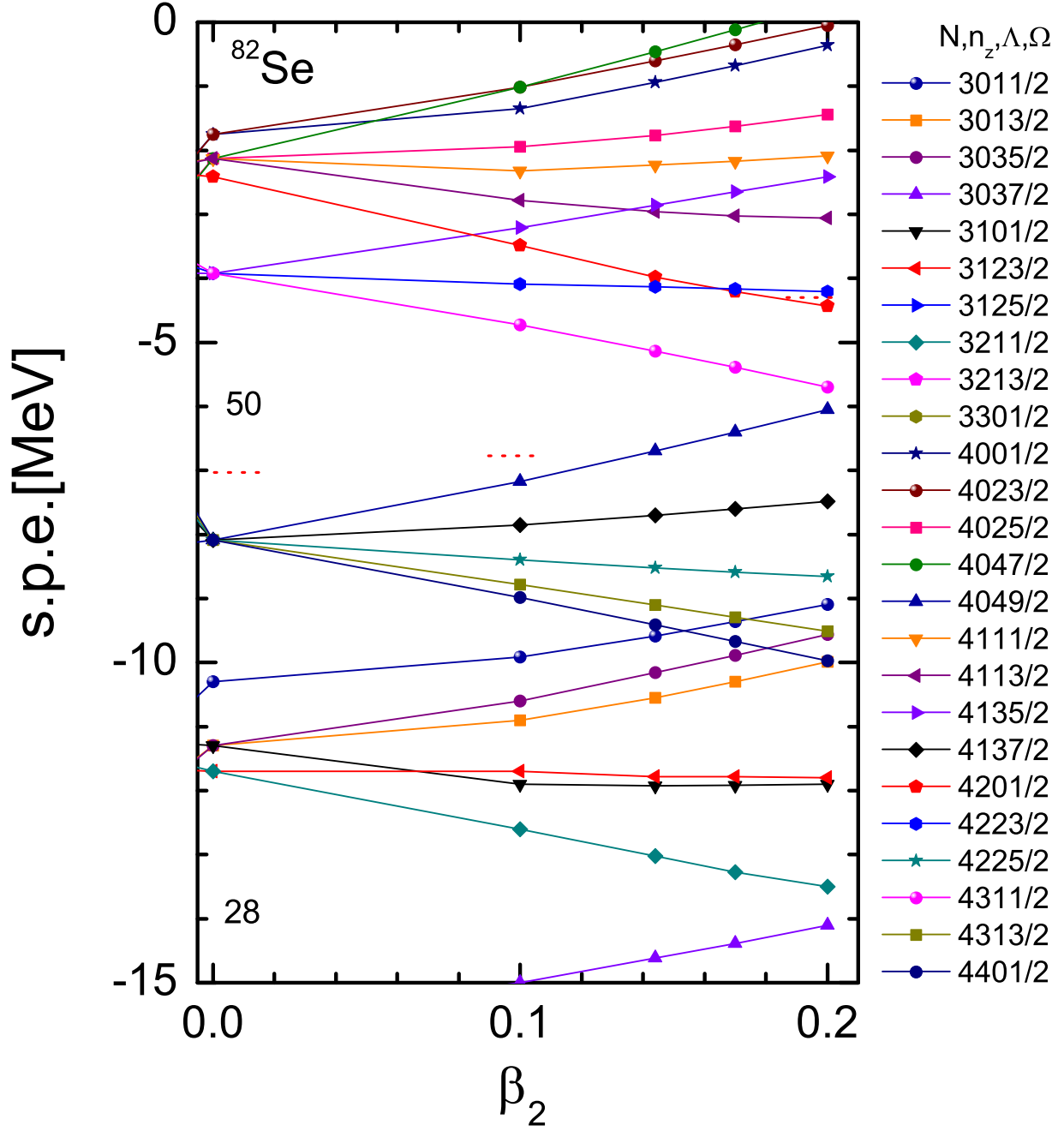}
\caption{(Color online) Single particle energies of $ ^{82}$Se as a function of the different deformation parameter $\beta_2 $. Red dotted lines denote the Fermi energy.
}
\label{fig3}
\end{figure}
\begin{figure}
\includegraphics[width=1.0\linewidth]{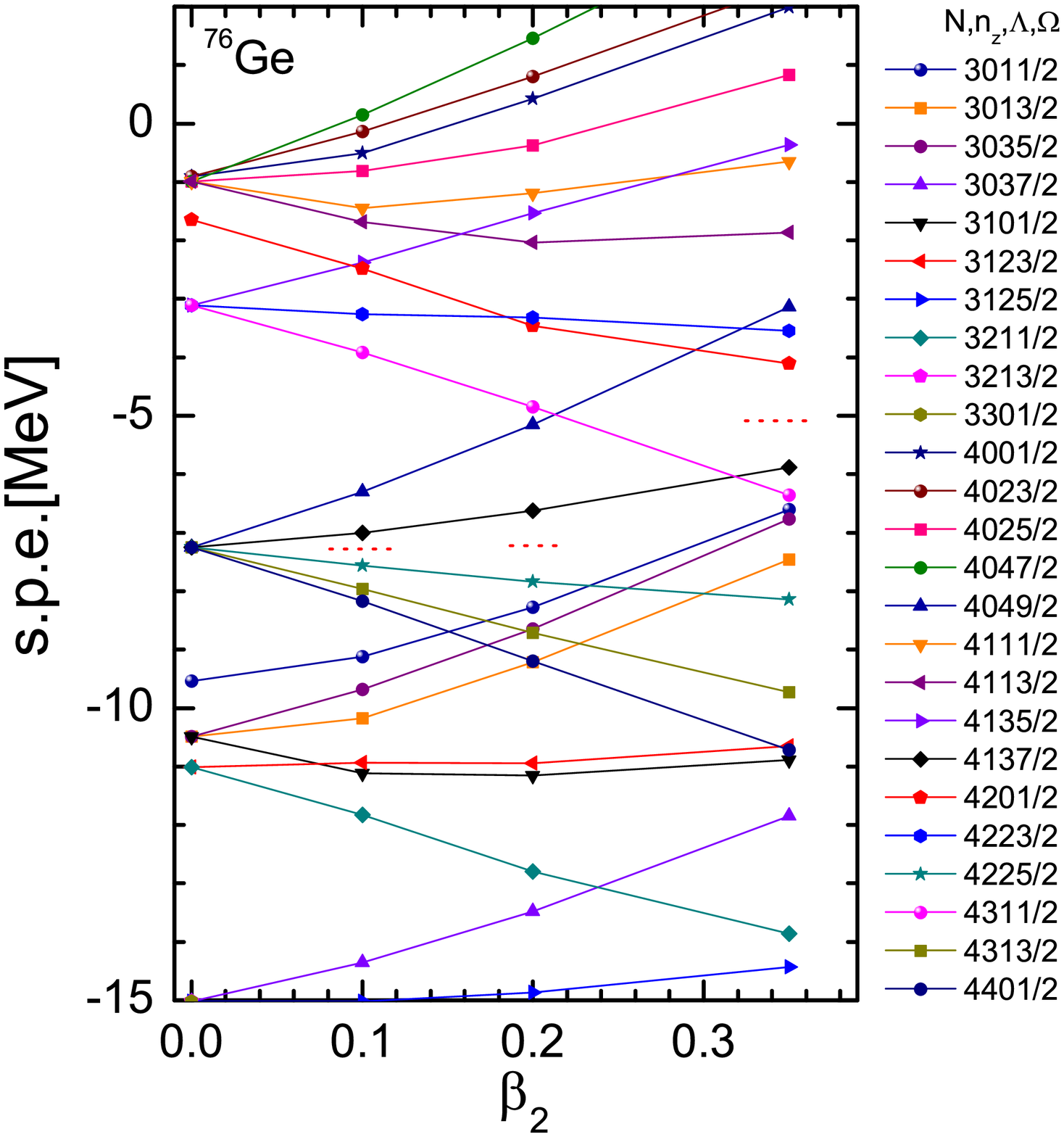}
\caption{(Color online) Same as Fig.3 for $ ^{76}$Ge
}
\label{fig4}
\end{figure}
We have taken $g_{ph}(g_{pp})=1.0(1.0)$ for the particle-hole (particle-particle) strength parameters for the two nuclei.
Since the nuclei considered here are expected to have large energy gaps between proton and neutron spaces,
in this work, we considered only the $nn$ and $pp$ pairing correlation although the formalism is presented generally.
For example, in the neutron-rich nuclei of importance in the r-process, the {\it np} pairing may not contribute so much.
But for the p-process, the {\it np} pairing could be more important than the neutron-rich nuclei because of the adjacent
energy gaps of protons and neutrons. The calculations for the neutron-deficient nuclei in p-process are in progress.

In the following, we show the GT strength distributions in terms of the $\beta_2$ parameter, as a function of excited energy of parent nuclei. Therefore, experimental data are presented by adding the empirical Q values from the measured data.
In Fig 5, the GT strength distribution, B(GT$^-$), on $^{82}$Se is presented as a
function of the excitation energy $E_{ex}$ w.r.t. the ground state of $^{82}$Se for $\beta_2 = 0 \sim \pm 0.2$.
Uppermost panel represents the experimental data deduced from the $^{82}$Se(p,n)$^{82}$Br reaction at 134.4MeV \cite{Madey89}. With the redistribution of the GT strengths, the GT data around $ 12$MeV are well reproduced at $\beta_2$ = 0.2 (see (c) panel in Fig.5). For a reference, the $\beta_2$ value from the relativistic mean field (RMF) is 0.133 \cite{Lala99}.
\begin{figure}
\includegraphics[width=0.8\linewidth]{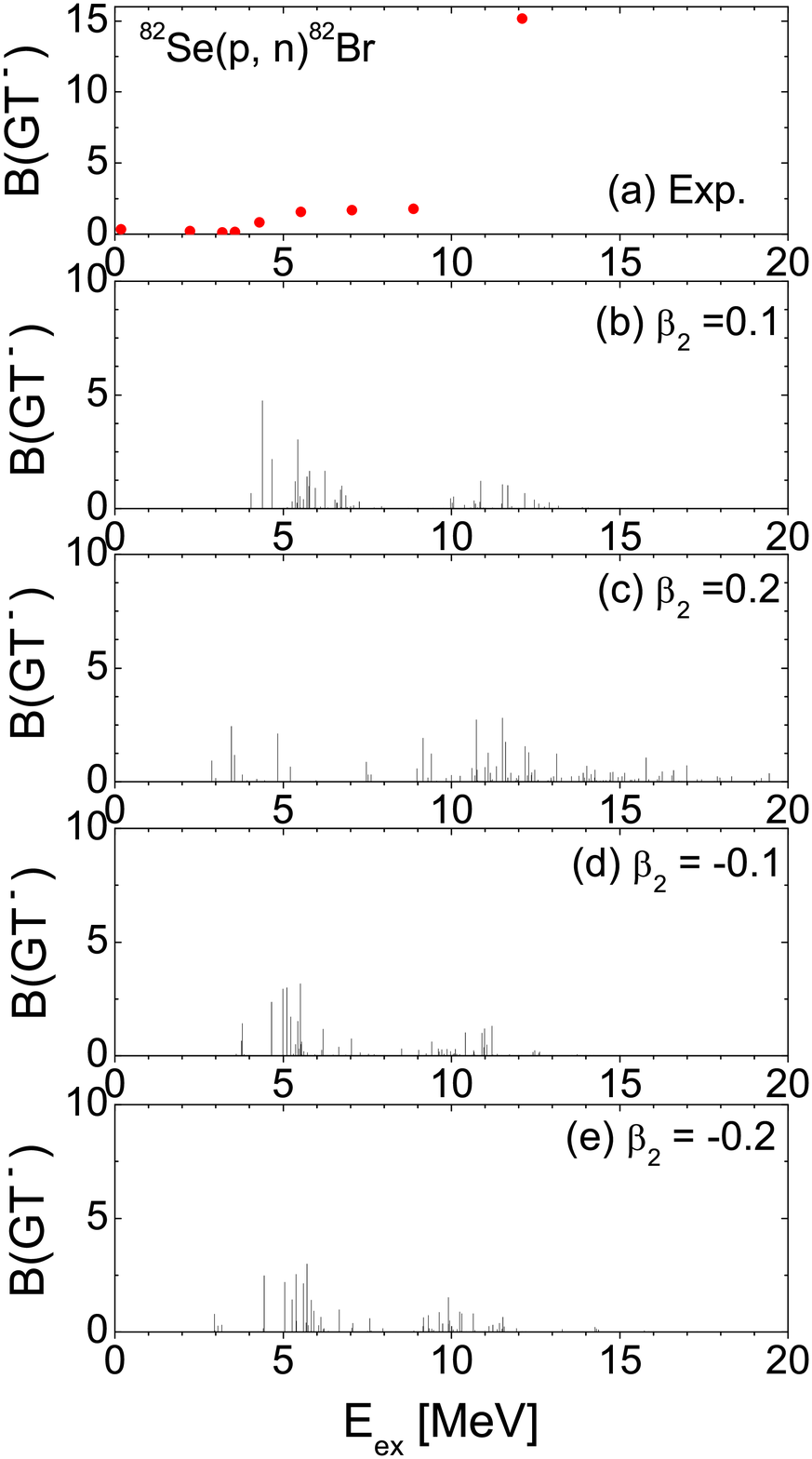}
\caption{(Color online) Gamow-Teller strength distributions B(GT$^{-}$)
on $ ^{82}$Se as a function of the excitation energy $E_{ex}$ w.r.t. the ground state of $^{82}$Se.
Experimental data denoted as filled (red) points in uppermost panel
are deduced from the $^{82}$Se(p,n) reaction at 134.4MeV \cite{Madey89}. In each panel, we indicate $\beta_2$.}
\label{fig5}
\end{figure}
\begin{figure}
\includegraphics[width=1.0\linewidth]{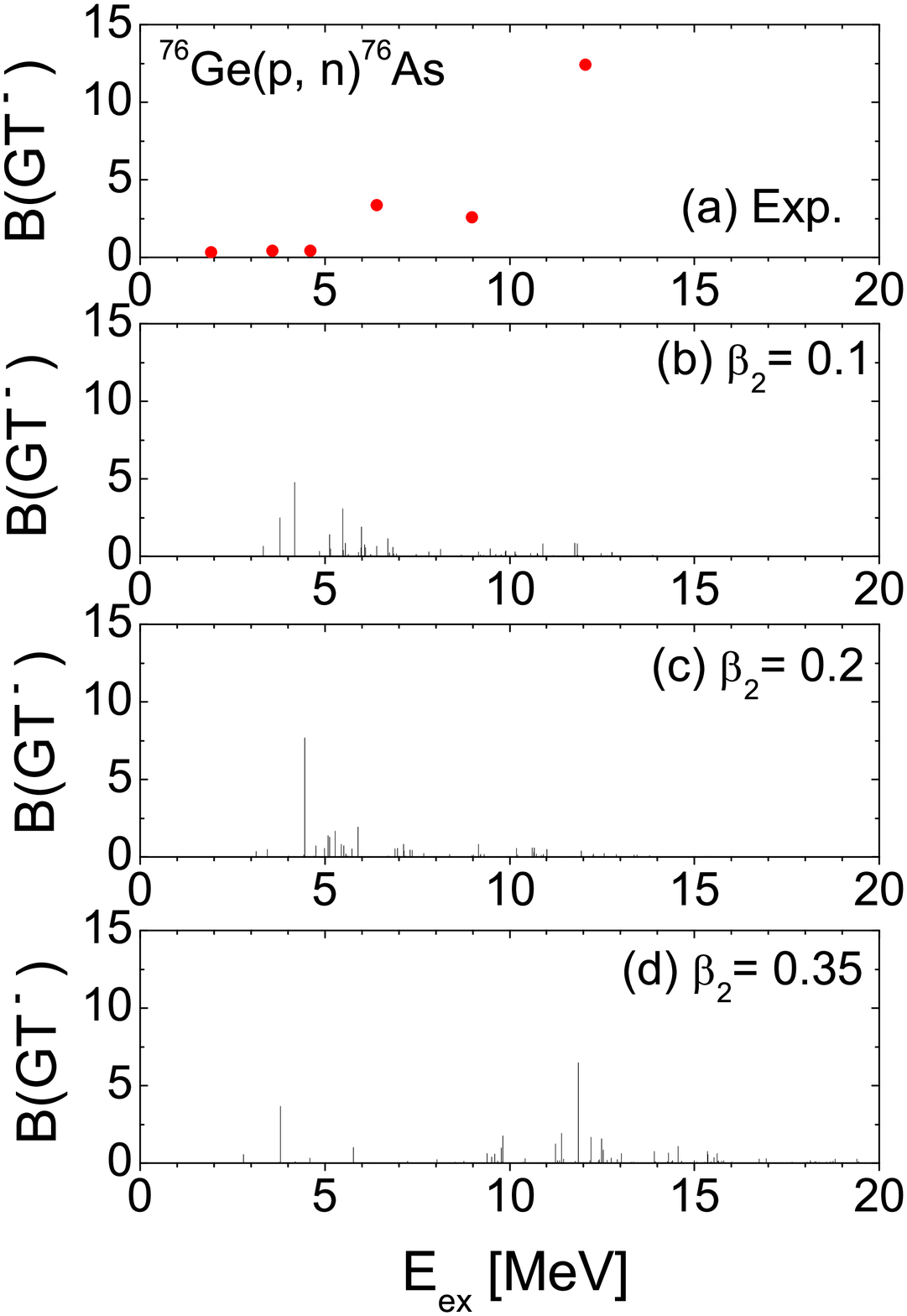}
\caption{(Color online) Same as for Fig.5 for $ ^{76}$Ge as a function of the excitation energy $E_{ex}$ w.r.t. the ground state of $^{76}$Ge. Experimental data denoted as filled (red) points in uppermost panel are deduced from the $^{76}$Ge(p,n) reaction at 134.4MeV
\cite{Madey89}.}
\label{fig6}
\end{figure}
Fig.6 shows B(GT$^-$) of $^{76}$Ge as a
function of the excitation energy $E_{ex}$ with prolate shapes, $\beta_2 = 0 \sim 0.35$.
Uppermost panel represents the experimental data from the $^{76}$Ge(p,n)$^{76}$As reaction at 134.4MeV \cite{Madey89},
which show a strong GT state peak around 12 MeV. The GT strength distributions
are widely scattered due to deformation similarly to Fig.5. In particular, results for the $\beta_2$ = 0.35,
in which the ISR I and II are almost satisfied, nicely reproduce the peak on the high-lying excited states. The value calculated by RMF is 0.157 \cite{Lala99}.

Both nuclei, $^{76}$Ge and $^{82}$Se, show a strong peak on high-lying GT states around 12 MeV. Our $\beta_2$ values, $\beta_2$=0.35 and 0.2 for $^{76}$Ge and $^{82}$Se, customized to the
ISR values, are shown to reproduce neatly the GT states.
Actually, these high-lying GT states turn out to be intimately associated with the smearing of some physical states around the Fermi surface in deformed nuclei.
\begin{figure}
\includegraphics[width=1.0\linewidth]{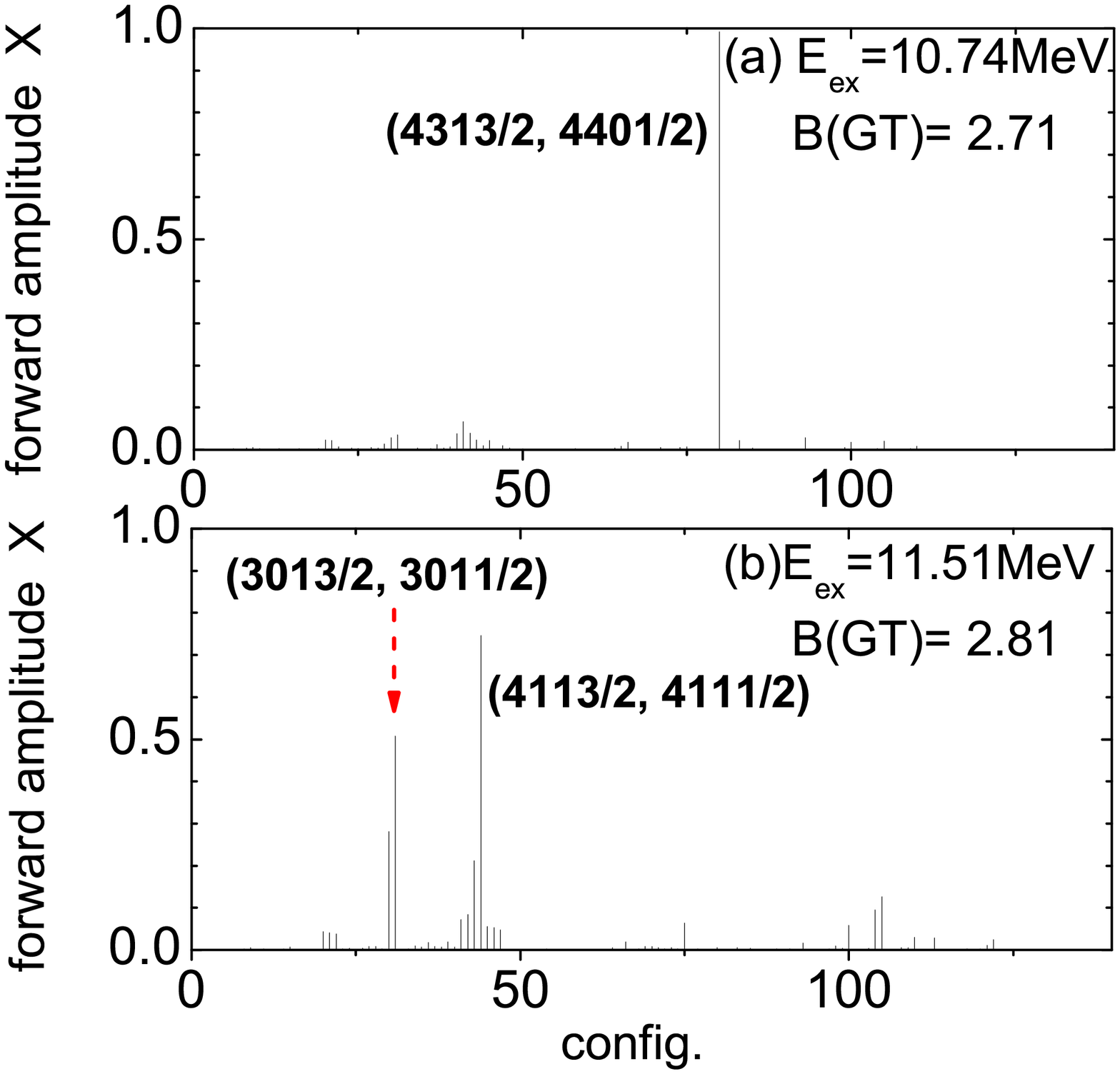}
\caption{Main transitions in the forward amplitudes $X$ for two dominant high-lying GT states ($E_{ex}$ = 10.74 and 11.51 MeV) w.r.t. the ground state of $^{82}$Se as a function of the configuration state for the panel (c) in Fig.5.}
\label{fig7}
\end{figure}
\begin{figure}
\includegraphics[width=1.0\linewidth]{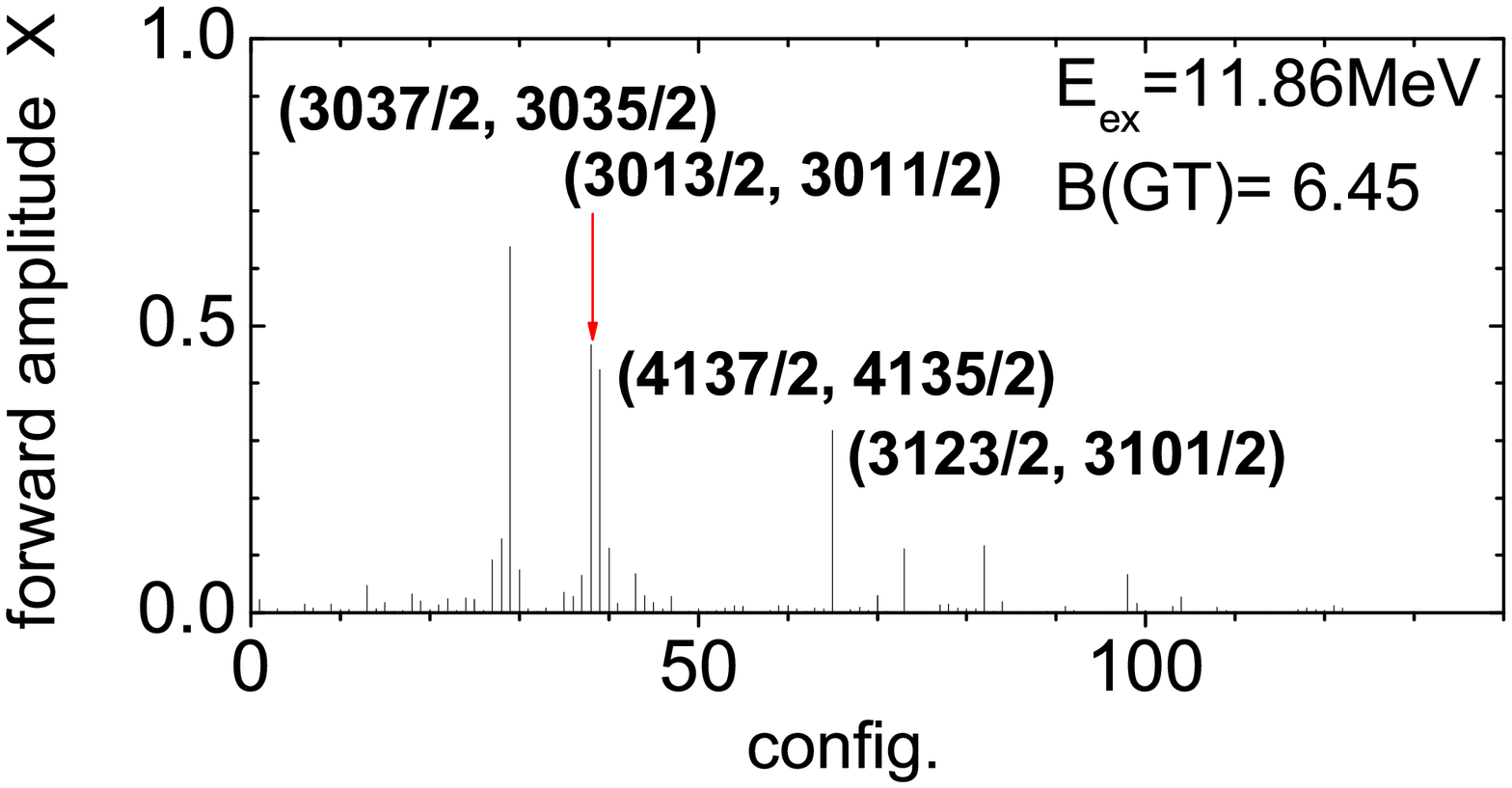}
\caption{ Same as Fig. 7 for $^{76}$Ge ($E_{ex}$ = 11.86 MeV) corresponding to the panel (d) in Fig.6}
\label{fig8}
\end{figure}
Figs.7 and 8 show the forward amplitudes $X$ of two excitation energies and one excitation energy, having the large B(GT) values in (c) and (d) of Figs.5 and 6, respectively.
For $^{82}$Se, the main high-lying GT transition come from (4 3 1 3/2)$\rightarrow$ (4 4 0 1/2), (3 0 1 3/2)$\rightarrow$ (3 0 1 1/2) and
(4 1 1 3/2)$\rightarrow$ (4 1 1 1/2) state transitions, while (3 0 3 7/2)$\rightarrow$(3 0 3 5/2), (3 0 1 3/2)$\rightarrow$ (3 0 1 1/2), (4 1 3 7/2)$\rightarrow$ (4 1 3 5/2) and (3 1 2 3/2)$\rightarrow$ (3 1 0 1/2) turn out to be main transitions for $^{76}$Ge.

In summary, to describe the single particle state in deformed basis, we used the deformed axially symmetric Woods-Saxon potential. We performed the
deformed BCS and deformed QRPA with a realistic two-body interaction calculated by {\it ab initio} G-matrix based on Bonn
potential. Results of the Gamow-Teller strength, B(GT), for $^{76}$Ge and $^{82}$Se show that the deformation effect leads to a fragmentation of the GT strength and reproduces the high-lying GT excitation deduced by higher energy projectiles, which result from the wide smearing by the increase of the Fermi surface energy due to the deformation. This work was supported by the National Research Foundation of Korea (C00020, 2012R1A1A3009733, 2011-0015467).

\end{document}